\documentstyle[12pt]{article}

\title{CONFORMAL TRANSFORMATIONS AND QUANTUM GRAVITY}
\author{FATIMAH SHOJAI\footnote{Email: FATIMAH@NETWARE2.IPM.AC.IR}$^{\ ,1}$
and ALI SHOJAI\footnote{Email: SHOJAI@NETWARE2.IPM.AC.IR}$^{\ ,2,1}$
and MEHDI GOLSHANI$^{3,1}$\\
$^1$Institute for Studies in Theoretical Physics and Mathematics,
\\P.O.Box 19395--5531, Tehran, IRAN.\\
$^2$Department of Physics, Tarbiat Moddarres University,
\\P.O.Box 14155-4838, Tehran, IRAN.\\
$^3$Department of Physics, Sharif University of Technology,
\\P.O.Box 11365--9161, Tehran, IRAN.}
\date{\today}
\begin{document}
\maketitle
\begin{abstract}
{\it Recently\cite{BQG}, it was shown that quantum effects of matter could be 
identified with the conformal degree of freedom of the space--time metric.
Accordingly, one can introduce quantum effects either by
making a scale transformation (i.e. changing the metric), or by making
a conformal transformation (i.e. changing all physical quantities).
These two ways are investigated and compared. Also, it is argued that,
the ultimate formulation of such a quantum gravity theory should be in
the framework of the scalar--tensor theories.}
\end{abstract}
\section{Quantum Effects, Scale, and Conformal Transformations}
In a previous paper\cite{BQG} it was shown that the application of an idea of
 de-Broglie leads to the fact that  the 
quantum effects of matter are equivalent to a specific conformal factor of the
 space--time metric. In the de-Broglie--Bohm relativistic quantum theory\cite{BOHM}, the
Klein--Gordon equation\footnote{A "bar" sign over any quantity indicates
 that it is expressed in terms of $\overline{g}_{\mu\nu}$ metric. This 
is introduced for later convenience.}
\begin{equation}
\stackrel{-}{\Box}\Psi+\frac{\overline{m}^2}{\hbar^2}\Psi=0
\end{equation}
is replaced with the quantum Hamilton--Jacobi equation:
\begin{equation}
\overline{g}^{\mu \nu}\overline{\nabla}_{\mu}S\overline{\nabla}_{\nu}S=\overline{\cal M}^2
\end{equation}
and the countinuty equation:
\begin{equation}
\overline{\nabla}_{\mu}(\overline{\rho}{\overline{\nabla}}^{\mu}S)=0
\end{equation} 
where all covariant derivatives are calculated with respect to the
 space--time metric $\overline{g}_{\mu \nu}$ and 
\begin{equation}
\Psi=\sqrt{\overline{\rho}}e^{iS/\hbar}
\end{equation}
and
\begin{equation}
\overline{\cal M}=\overline{m}(1+\overline{Q})^{1/2}
\end{equation}
\begin{equation}
\overline{Q}=\alpha\frac{\stackrel{-}{\Box}
\sqrt{\overline{\rho}}}{\sqrt{\overline{\rho}}};\ \ \ \ \ \ 
\alpha=\frac{\hbar^2}{\overline{m}^2}
\end {equation}
It muse be noted that $\overline{{\cal M}}$ is a variable mass, originating
 from the matter quantum potential $\overline{Q}$.

If the above metric is written as conformally transformed of another metric
($g_{\mu \nu}$) which 
\begin{equation}
\overline{g}_{\mu \nu}={\phi}^{-1}{g}_{\mu \nu}
\end{equation}
where
\begin{equation}
{\phi}^{-1}=1+\overline{Q}
\end{equation}
Then the Hamilton--Jacobi equation in terms of this metric is:
\begin{equation}
{g}^{\mu \nu}{\nabla}_{\mu }S{\nabla}_{\nu}S=\overline{m}^2
\end{equation}
This means that quantum effects has been removed in this new metric.
In other words, in terms of $\overline{g}_{\mu \nu}$ (the background metric) the
quantum effects are included in the variable mass $\overline{\cal M}$, while
in terms of $g_{\mu \nu}$ metric some parts of the curvature of the space--time
represents the quantum effects. This new metric is called the physical metric  
because it has physical interpretation. For example the singularities of FRW
universe would be removed in $g_{\mu \nu}$\cite {BQG}.

In our previous work \cite{BQG}, the conformal transformation was applied only
 to the space--time metric.
Other quantities like mass, density and so on were assumed to posses no
 transformation. This is because
the above conformal transformation which incorporates the quantum effects
 of matter into a specific conformal 
factor, is in fact a scale transformation guessed from the Hamilton--Jacobi
 equation
 (2). As the conformal transformation is more general than scale
 transformation which is used in \cite{BQG}, it seems preferable
 to make a conformal transformation, in
 which all physical quantities are transformed, 
instead of making only a scale transformation.

Now, applying the conformal transformation given by the equation
 (7), we have:
\begin{equation}
\overline{m}={\phi}^{1/2}m
\end{equation}
and
\begin{equation}
\overline{\rho}={\phi}^{3/2}\rho
\end{equation}
Therefore the equation (2) reads:
\begin {equation}
\nabla_{\mu}S\nabla^{\mu}S={\phi}^2 m^2(1+\overline{Q})
\end{equation}
instead of (9).

In order to have no quantum effects, when physics is expressed in terms
 of the physical metric, it is 
necessary to set:
\begin{equation}
\phi^{-2}=1+\overline{Q}
\end{equation} 
This equation determines the conformal degree of freedom of the space--time
metric. 

With the aid of the guiding equation in the physical metric 
(${P}^{\mu}={m}{u}^{\mu}=\nabla^\mu S$) and the equation (12),
the equation of motion takes the form:
\begin{equation}
\frac{d{u}_{\mu}}{d{\tau}}-\frac{1}{2}\left ( \partial_{\mu} 
{g}_{\alpha\beta}\right ){u}^{\alpha}{u}^{\beta}=0
\end{equation}
which is the geodesic equation for a free particle. But with respect to the 
background metric ($\overline{g}_{\mu \nu}$), the particle is affected by quantum force 
and therefore it dosen't move on the geodesic. In this case from the 
guiding formula ($\overline{P}^{\mu}=\overline{\cal M}\overline{u}^{\mu}=
\overline{\nabla}^\mu S$)
and the equation (2), we have
\begin{equation}
\frac{d\overline{u}_{\mu}}{d\overline{\tau}}-\frac{1}{2}\left ( \partial_{\mu} 
\overline{g}_{\alpha\beta}\right )\overline{u}^{\alpha}\overline{u}^{\beta}=\frac{1}{\overline{\cal M}}\left (
\overline{g}_{\mu\nu}-\overline{u}_{\mu}\overline{u}_{\nu}\right )\overline{\nabla}^{\nu}\overline{\cal M}
\end{equation}
where the expression on the right hand side is the quantum force\cite{BOHM}.

It can be shown that these two geodesic equations (14,15), as expected, are
conformal transformations of each other. Using the conformal 
transformation, equation (14) reads: 
\begin{equation}
\frac{d\overline{u}_{\mu}}{d\overline{\tau}}-\frac{1}{2}\left ( \partial_{\mu} 
\overline{g}_{\alpha\beta}\right )\overline{u}^{\alpha}\overline{u}^{\beta}=
-\frac{1}{2}\phi^{-1/2}\overline{u}_{\mu}\frac{d\phi}{d\overline{\tau}}-\frac{1}{2}
\overline{u}^{\alpha}\overline{u}^{\beta}\overline{g}_{\alpha\beta}\frac{\partial_{\mu}\phi}{\phi}
\end{equation}
Since $d/d\overline{\tau}=\overline{u}_{\alpha}\overline{\nabla}^{\alpha}$ and 
$\phi=\overline{m}/\overline{\cal M}$, a simple calculation shows that the right
hand side of the above equation is the quantum force.

Here, It must be noted that the conformal factor is obtained only from
one of the equations of motion, i.e. the Hamilton--Jacobi equation. The continouty
equation in the physical metric is:
\begin{equation}
{\nabla}_{\mu}\left ( {\rho}{\nabla}^{\mu}S\right )=0
\end{equation}
So
\begin{equation}
\overline{\nabla}_{\mu}\left (\overline{ \rho}\overline{\nabla}^{\mu}S
\right )=-\frac{{\overline{\nabla}}_{\mu}\phi}{2\phi}
\overline{\rho}\overline{\nabla}^{\mu}S
\end{equation}
The right hand side of this expression is of the order of $\hbar^2$  and higher\cite{BQG}.

The conformal transformation used to show the equivalence between quantum 
effects and cuavature of space--time has many applications. It is possible 
to discuss the physical properties of the conformal factor (13) as it was
done in our earlier paper\cite{BQG} about the scale
 factor (8). Correspondingly, one can start from gravity--matter action
 and then proceed to eliminate the matter 
quantum potential and transform metric and physical quantities conformally.
Therefore, the quantum gravity action is:
\begin{equation}
{\cal A}=\int d^4x \sqrt{-\overline{g}}\left \{ \phi 
\overline{\cal R}-\frac{3}{2}
\frac{\overline{\nabla}_{\mu}\phi\overline{\nabla}^{\mu}
\phi}{\phi}+2\overline{\kappa}\left (
\phi^3\frac{\overline{\rho}}{\overline{m}}\overline{\nabla}_{\mu}S
\overline{\nabla}^{\mu}S
-\phi\overline{\rho}\ \overline{m}\right )
+\Lambda \left ( \phi-(1+\overline{Q})^{-1/2}\right )\right \}
\end{equation}
where $\Lambda$ is a lagrangian multiplier introduced in order to fix
the conformal factor.
Variation of this action with respect to $\overline{g}_{\mu \nu}$, $\phi$
, $\overline{\rho}$, $S$ and $\Lambda$ leads to:
\begin{enumerate}
\item The equation of motion of $\phi$:
\begin{equation}
\overline{\cal R}-\frac{3}{2}\frac{\overline{\nabla}_{\mu}\phi
\overline{\nabla}^{\mu}\phi}{\phi^2}
+3\frac{\stackrel{-}{\Box}\phi}{\phi}+\Lambda+6\overline{\kappa}
\phi^2\frac{\overline{\rho}}{\overline{m}}\overline{\nabla}_{\mu}S
\overline{\nabla}^{\mu}S-2\overline{\kappa}\ \overline{\rho}\ \overline{m}=0
\end{equation}
\item The continouty equation:
\begin{equation}
\overline{\nabla}_\mu\left ( \overline{\rho} \phi^3 
\overline{\nabla}^{\mu}S\right )=0
\end{equation}
\item The equation of motion of particles:
\begin{equation}
\frac{4\overline{\kappa}}{\overline{m}}\phi^3\sqrt{\overline{\rho}}\ 
\overline{\nabla}_{\mu}S\overline{\nabla}^{\mu}S-4\overline{\kappa}\ \overline{m}
\sqrt{\overline{\rho}}\phi-\frac{\Lambda}{2}(1+\overline{Q})^{-3/2}
\frac{\overline{Q}}{\sqrt{\overline{\rho}}}
+\frac{\Lambda\alpha}{2}\stackrel{-}{\Box}\left (
\frac{(1+\overline{Q})^{-3/2}}{\sqrt{\overline{\rho}}}\right )
=0
\end{equation}
\item The modified Einstien equations for $\overline{g}_{\mu \nu}$:
\[ \overline{\cal G}^{\mu\nu}-\frac{1}{2}\Lambda \overline{g}^{\mu\nu}=
\frac{1}{\phi}
[\overline{g}^{\mu\nu}\stackrel{-}{\Box}-\overline{\nabla}^\mu
\overline{\nabla}^\nu ]\phi+\frac{3}{2}\left (
\frac{\overline{\nabla}^\mu\phi\overline{\nabla}^\nu\phi}{\phi^2}-\frac{1}{2}
\frac{\overline{\nabla}^\alpha\phi\overline{\nabla}_\alpha\phi}{\phi^2}
\overline{g}^{\mu\nu}\right )\]
\[ +\frac{\overline{\kappa}\ \overline{\rho}}{\overline{m}}\phi^2
\overline{\nabla}^\alpha S\overline{\nabla}_\alpha S \overline{g}^{\mu\nu}
-\frac{2\overline{\kappa}}{\overline{m}}\overline{\rho}\phi^2
\overline{\nabla}^\mu S \overline{\nabla}^\nu S -\overline{\kappa}
\ \overline{m}\ \overline{\rho}\ 
\overline{g}^{\mu\nu}-\frac{1}{2}\frac{\Lambda}{\phi}(1+\overline{Q})^{-1/2}
\overline{g}^{\mu\nu} \]
\[ -\frac{\Lambda}{4}(1+\overline{Q})^{-3/2}\overline{Q}
\overline{g}^{\mu\nu}-\frac{\alpha}{4\phi}
\overline{g}^{\mu\nu}\overline{\nabla}_\alpha\sqrt{\overline{\rho}}\ 
\overline{\nabla}^\alpha\left  ( \Lambda \phi
\frac{(1+\overline{Q})^{-3/2}}{\sqrt{\overline{\rho}}}\right )\]
\begin{equation}
+\frac{\alpha}{4\phi}\overline{\nabla}^\mu\sqrt{\overline{\rho}}\ 
\overline{\nabla}^\nu\left ( \Lambda \phi
\frac{(1+\overline{Q})^{-3/2}}{\sqrt{\overline{\rho}}}\right )
+\frac{\alpha}{4\phi}\overline{\nabla}^\nu\sqrt{\overline{\rho}}\ 
\overline{\nabla}^\mu\left ( \Lambda \phi
\frac{(1+\overline{Q})^{-3/2}}{\sqrt{\overline{\rho}}}\right )
\end{equation}
\item The constraint equation:
\begin{equation}
\phi^{-2}=1+\overline{Q}
\end{equation}
\end{enumerate}
Then, the equations of motion can be solved, leading to the background 
metric and other physical quantities. 
\section{Concluding Remarks}
Recently\cite{BQG}, it was shown that quantum effects could be contained
in the conformal factor of the space--time metric. In this paper, the difference
between introducing the quantum effects by conformal transformation or
scale transformation are discussed. No essential difference is found. But, some
 points must be noted here.
The first important point is about the geodesic equation (14,15). In the 
background metric, this equation resembles the geodesic equation in
Brans--Dicke theory. Consideration of the matter quantum effects, leads 
to the physical metric in which a particle moves on the geodesic of
Branse--Dicke theory written in Einstein guage. This point supports 
the suggestion that the discussion of quantum gravity requires a
scalar--tensor theory. Previously this was suggested when disscusing
 Bohmian quantum gravity \cite{QGC}.

Secondly, since as the dimensional coupling between matter and gravity is 
resulted from the breaking of the conformal invariance, this formulation 
shows that {\it the conformal frame is fixed by the distribution of matter
at quantum level. Thus, the Bohmian quantum theory singles out
the prefered frame\/}.

In the present paper and the earlier one \cite{BQG} we have used lagrangian
multiplier, in order to fix the conformal factor by quantum potential.
We suggested now that it must be possible to write a scalar--tensor theory
which automatically leads to the correct equations of motion. This has the 
advantage that the conformal factor would be fixed by the equations of 
motion, and not by introducing a lagrangian multiplier by hand.
We shall elaborate on this point in a forthcoming paper\cite{STQ}.

\end{document}